\documentclass{article}
\usepackage{amssymb}
\usepackage{amsmath}
\usepackage{graphicx}

\begin{document}

\begin{center}
\textbf{UPPER\ LIMIT\ ON\ THE CENTRAL\ DENSITY OF DARK MATTER IN THE
EDDINGTON INSPIRED BORN-INFELD (EiBI) GRAVITY}

\bigskip

Ramil Izmailov$^{1,a}$, Alexander A. Potapov$^{2,b}$, Alexander I. Filippov$%
^{2,c}$,

Mithun Ghosh$^{3,d}$ and Kamal K. Nandi$^{1,2,3,e}$

$\bigskip $

$^{1}$Zel'dovich International Center for Astrophysics, M. Akmullah Bashkir State
Pedagogical University, Ufa 450000, RB, Russia

$^{2}$Department of Physics \& Astronomy, Bashkir State University,
Sterlitamak Branch, Sterlitamak 453103, RB, Russia \\[0pt]

$^{3}$ Department of Mathematics, University of North Bengal, Siliguri
734013, WB, India \\[0pt]

\bigskip

$^{a}$Email: izmailov.ramil@gmail.com

$^{b}$Email: potapovaa@mail.ru

$^{c}$Email: filippovai@rambler.ru

$^{d}$Email: ghoshmithun123@gmail.com

$^{e}$Email:kamalnandi1952@yahoo.co.in

\bigskip
\end{center}

PACS\ number(s): 95.35+d, 97.20.Vs, 04.50.1h

\begin{center}
\textbf{Abstract}
\end{center}

We investigate the stability of circular material orbits in the analytic
galactic metric recently derived by Harko \textit{et al.} (2014). It turns
out that stability depends more strongly on the dark matter central density $%
\rho _{0}$ than on other parameters of the solution. This property then
yields an upper limit on $\rho _{0}$ for each individual galaxy, which we
call here $\rho _{0}^{\text{upper}}$, such that stable circular orbits are
possible \textit{only} when the constraint $\rho _{0}\leq \rho _{0}^{\text{%
upper}}$ is satisfied. This is our new result. To approximately quantify the
upper limit, we consider as a familiar example our Milky Way galaxy that has
a projected dark matter radius $R_{\text{DM}}\sim 180$ kpc and find that $%
\rho _{0}^{\text{upper}}\sim 2.37\times 10^{11}$ $M_{\odot }$kpc$^{-3}$.
This limit turns out to be about four orders of magnitude larger than the
latest data on central density $\rho _{0}$ arising from the fit to the
Navarro-Frenk-White (NFW) and Burkert density profiles. Such consistency
indicates that the EiBI solution could qualify as yet another viable
alternative model for dark matter.
\bigskip

\begin{center}
---------------------------------------
\end{center}

The issue of dark matter is at the centerstage of modern astrophysics. Several theoretical models and cosmological scenarios for dark matter exist in the
literature[1], among which the possibility of perfect fluid dark matter within
the framework of general relativity has already been explored in the
literature [2,3]. A similar possibility has been recently investigated
within the framework of the Eddington inspired Born-Infeld (EiBI) theory by
Harko \textit{et al.}[4].

Using a tangential velocity profile [5] giving
flat rotation curves at large distances and setting the cosmological
constant $\Lambda$ to zero, they obtained, in the Newtonian approximation,
a new galactic metric and theoretically explored its gravitational
properties including the dark matter density distribution.

However, the numerical values of the crucial parameter $\kappa $ 
(denoted by $\kappa =2R_{\text{DM}}^{2}/\pi^{2}$) or equivalently the dark matter radius $R_{\text{DM}}$ cannot be determined from the theory alone $-$ it has to be supplied by
the observed data. This $R_{\text{DM}}$ is expected to make its appearance
also in the stability of orbits.

The purpose of the present Letter is to investigate the stability of
circular material orbits in the EiBI galactic metric [4], which is by no
means obvious from the metric itself. It turns out that the stability is
exclusively sensitive to the variation of the dark matter central density $\rho _{0}$ leading to an upper limit, which we call here $\rho _{0}^{\text{upper}}$, such that stable circular orbits in the EiBI are possible only
when the constraint $\rho _{0}\leq \rho _{0}^{\text{upper}}$ is satisfied.
We shall illustrate the inequality by the example of our own Milky Way
galaxy, which shows that $\rho _{0}^{\text{upper}}$ is about four orders of
magnitude \textit{larger} than the latest data on $\rho _{0}$ arising from
the fit to NFW or Burkert profiles. We shall take units such that $G=1$, $c=1 $, 
unless otherwise specified.

The salient features of the EiBI dark matter model are as follows: The
action is%
\begin{equation}
S_{\text{EiBI}}=\frac{1}{8\pi \kappa }\int d^{4}x\left[ \sqrt{-\left\vert
g_{\mu \nu }+8\pi \kappa R_{\mu \nu }\right\vert }-\lambda \sqrt{-g}\right]
+S_{\text{matter}},
\end{equation}%
where $\lambda $ is a dimensionless parameter and $\kappa $ is a parameter
with inverse dimension to that of the cosmological constant $\Lambda $. In
the limit $\kappa \rightarrow 0$, the Hilbert-Einstein action is recovered
with $\lambda =8\pi \kappa \Lambda +1$, where $\Lambda $ is the cosmological
constant. Harko \textit{et al.} [4] developed spherically symmetric solution
assuming $\lambda =1\Rightarrow \Lambda =0$ so that $\kappa $ can have
nonzero values. The description of the physical behavior of various
cosmological and stellar scenarios was assumed to be controlled by the
single parameter $\kappa $. The galactic halo is assumed to be filled with
perfect fluid dark matter with energy-momentum tensor $T^{\mu \nu }=pg^{\mu
\nu }+(p+\rho )U^{\mu }U^{\nu }$, $g_{\mu \nu }U^{\mu }U^{\nu }=-1$. There
are two metrics in the EiBI theory, the physical metric $g_{\mu \nu }$ and
the auxiliary metric $q_{\mu \nu }$ and the tangential velocity profile is
taken as [5]%
\begin{equation}
v_{\text{tg}}^{2}=v_{\infty }^{2}\frac{(r/r_{\text{opt}})^{2}}{(r/r_{\text{%
opt}})^{2}+r_{0}^{2}},
\end{equation}%
where $r$ is the standard radial coordinate (as defined by \textit{Eq.(9)}
in [4]), $r_{\text{opt}}$ is the optical radius containing 83\% of the
galactic luminosity. The parameter $r_{0}$ is defined as the ratio of the
halo core radius and $r_{\text{opt}}$, and $v_{\infty }$ is the asymptotic
constant velocity. Under the Newtonian approximations that the pressure $%
p\simeq 0$, $8\pi \kappa \rho \ll 1$, and $(r/r_{\text{opt}})^{2}\gg
r_{0}^{2}$, the EiBI field equations yield the Lane-Emden equation with
polytropic index $n=1$, which has an exact nonsingular solution for dark
matter density as [4] 
\begin{equation}
\rho ^{(0)}(r)=K\left[ \frac{\sin \left( r\sqrt{\frac{2}{\kappa }}\right) }{r%
\sqrt{\frac{2}{\kappa }}}\right] ,
\end{equation}%
where $\rho ^{(0)}(0)=K=\rho _{0}$ is the central density.\footnote{%
The solution (3) of the Lane-Emden equation and its connection to dark
matter were first pointed out in Refs.[26-28]. The same density profile is
used also in the context of the Bose-Einstein Condensate (BEC) simulation of
dark matter [29].} Assuming that the halo has a sharp boundary $R_{\text{DM}}
$, where the density vanishes such that $\rho ^{(0)}(R_{\text{DM}})=0$, one
has%
\begin{equation}
R_{\text{DM}}=\pi \sqrt{\frac{\kappa }{2}}.
\end{equation}%
Thus, the mass profile of the dark matter is 
\begin{equation}
M(r)=4\pi \int_{0}^{r}\rho ^{(0)}(r)r^{2}dr=\frac{4R_{\text{DM}}^{3}}{\pi
^{2}}\rho _{0}\left[ \sin (\overline{r})-\overline{r}\cos (\overline{r})%
\right] ,
\end{equation}%
where the dimensionless quantity $\overline{r}=\pi r/R_{\text{DM}}$. The
average velocity dispersion of dark matter particles in the constant
velocity region in the present model is [4]%
\begin{equation}
\sigma ^{2}=\left\langle \overrightarrow{v}^{2}\right\rangle /3=v_{\text{tg}%
}^{2}\left[ 1-\frac{\overline{r}}{\sin (\overline{r})}\text{Ci}\left( 
\overline{r}\right) \right] ,
\end{equation}%
where Ci$\left( z\right) =-\int_{z}^{\infty }\frac{\cos (t)}{t}dt$ is the
cosine integral function.

The approximate physical metric has been derived as [4] 
\begin{equation}
d\tau ^{2}=-B(\overline{r})dt^{2}+A(\overline{r})d\overline{r}^{2}+\overline{%
r}^{2}C(\overline{r})(d\theta ^{2}+\sin ^{2}\theta d\phi ^{2}),
\end{equation}%
\begin{equation}
B(\overline{r})=e^{\nu _{0}}\left[ \left( \frac{R_{\text{DM}}}{\pi r_{\text{%
opt}}}\right) ^{2}\overline{r}^{2}+r_{0}^{2}\right] ^{v_{\infty }^{2}},
\end{equation}%
\begin{equation}
A(\overline{r})=\left( \frac{R_{\text{DM}}}{\pi }\right) ^{2}\frac{1}{\left[
1-\frac{\overline{\rho }_{0}}{\overline{r}}\sin (\overline{r})+\overline{%
\rho }_{0}\cos (\overline{r})\right] \left[ 1-\frac{\overline{\rho }_{0}}{%
\overline{r}}\sin (\overline{r})\right] },
\end{equation}

\begin{equation}
C(\overline{r})=\left( \frac{R_{\text{DM}}}{\pi }\right) ^{2}\left[ 1-\frac{%
\overline{\rho }_{0}}{\overline{r}}\sin (\overline{r})\right] ,
\end{equation}%
where $e^{\nu _{0}}$ is an arbitrary constant of integration (which we set
to unity) and the dimensionless quantity $\overline{\rho }_{0}=\frac{8\rho
_{0}R_{\text{DM}}^{2}}{\pi }$. Note that the surface area of a sphere at the
boundary of dark matter halo defined by $\overline{r}=\pi $, has the value $%
S=4\pi \overline{r}^{2}C(\overline{r})=4\pi \overline{r}^{2}\left( \frac{R_{%
\text{DM}}}{\pi }\right) ^{2}=4\pi R_{\text{DM}}^{2}$, which is just the
spherical surface area in "standard coordinates". Thus the dark matter
radius $R_{\text{DM}}$ can be identified with standard coordinate radius.

To analyze the stability of circular orbits, one needs to analyze the second
order derivative of the concerned potential, which we wish to do here. To
find the potential $V$, note that the four velocity $U^{\alpha }=\frac{%
dx^{\sigma }}{d\tau }$ of a test particle of rest mass $m_{0}$ moving in the
halo (restricting ourselves to $\theta =\pi /2$) follows the equation $%
g_{\nu \sigma }U^{\nu }U^{\sigma }=-m_{0}^{2}$ that can be cast into a
Newtonian form in the dimensionless radial variable $\overline{r}$ ($=\pi
r/R_{\text{DM}}$) as 
\begin{equation}
\left( \frac{d\overline{r}}{d\tau }\right) ^{2}=E^{2}+V(\overline{r})
\end{equation}%
which gives, for the metric Eqs.(7)-(10), the potential 
\begin{equation}
V(\overline{r})=\left[ E^{2}\left\{ \frac{1}{AB}-1\right\} -\frac{L^{2}}{AC%
\overline{r}^{2}}-\frac{1}{A}\right]
\end{equation}%
\begin{equation}
E=\frac{U_{0}}{m_{0}},L=\frac{U_{3}}{m_{0}},
\end{equation}%
where the constants $E$ and $L$, respectively, are the conserved
relativistic energy and angular momentum per unit mass of the test particle.
Circular orbits at any arbitrary radius are defined by $\overline{r}=%
\overline{R}=$ constant, so that $\frac{d\overline{r}}{d\tau }\mid _{%
\overline{r}=\overline{R}}=0$ and, additionally, $\frac{dV}{d\overline{r}}%
\mid _{\overline{r}=\overline{R}}=0$. From these two conditions follow the
conserved quantities: 
\begin{equation}
L^{2}=\frac{X}{Z}
\end{equation}%
and using it in $V(\overline{R})=-E^{2}$, we get 
\begin{equation}
E^{2}=\frac{Y}{Z},
\end{equation}%
where%
\begin{equation}
X\equiv -\kappa ^{2}\overline{R}^{3}v_{\infty }^{2}(\overline{R}-\overline{%
\rho }_{0}\sin \overline{R})^{2}
\end{equation}%
\begin{equation}
Y\equiv \left( \kappa \overline{R}^{2}+2r_{0}^{2}r_{\text{opt}}^{2}\right)
\left( r_{0}^{2}+\frac{\kappa \overline{R}^{2}}{2r_{\text{opt}}^{2}}\right)
^{v_{\infty }^{2}}\left( \overline{\rho }_{0}\overline{R}\cos \overline{R}+%
\overline{\rho }_{0}\sin \overline{R}-2\overline{R}\right)
\end{equation}%
\begin{eqnarray}
Z &\equiv &\left\{ \kappa \overline{R}^{2}\left( 1-2v_{\infty }^{2}\right)
+2r_{0}^{2}r_{\text{opt}}^{2}\right\} \overline{\rho }_{0}\sin \overline{R}%
+\left( \kappa \overline{R}^{2}+2r_{0}^{2}r_{\text{opt}}^{2}\right) 
\overline{\rho }_{0}\overline{R}\cos \overline{R}  \nonumber \\
&&-4\overline{R}r_{0}^{2}r_{\text{opt}}^{2}-2\kappa \overline{R}^{3}\left(
1-v_{\infty }^{2}\right) .
\end{eqnarray}

Putting the expressions for $L^{2}$ and $E^{2}$ in Eq.(12), we find the
complete expression for $V$. The orbits will be stable if $V^{\prime \prime
}\equiv $ $\frac{d^{2}V}{d\overline{r}^{2}}\mid _{\overline{r}=\overline{R}%
}<0$ and unstable if $V^{\prime \prime }>0$. The expression for $V^{\prime
\prime }$ is 
\begin{eqnarray}
&&V^{\prime \prime }\left( \overline{R};\kappa ,\overline{\rho }%
_{0},r_{0},r_{\text{opt}},v_{\infty }\right)  \nonumber \\
&=&\left[ \frac{2v_{\infty }^{2}\left( \overline{R}+\overline{\rho }_{0}%
\overline{R}\cos \overline{R}-\overline{\rho }_{0}\sin \overline{R}\right) }{%
\overline{R}^{2}\left( \kappa \overline{R}^{2}+2r_{0}^{2}r_{\text{opt}%
}^{2}\right) Z}\right] \times  \nonumber \\
&&\left[ 32\overline{R}^{2}r_{0}^{2}r_{\text{opt}}^{2}+8\kappa \overline{R}%
^{4}\left( 1-v_{\infty }^{2}\right) +6\overline{\rho }_{0}^{2}r_{0}^{2}r_{%
\text{opt}}^{2}\left( 1+\overline{R}^{2}\right) \right.  \nonumber \\
&&\left. +\kappa \overline{\rho }_{0}^{2}\overline{R}^{2}\left( 1-2v_{\infty
}^{2}+3\overline{R}^{2}\right) -2\overline{R}^{2}\left\{ 14r_{0}^{2}r_{\text{%
opt}}^{2}+\kappa \overline{R}^{2}\left( 5-2v_{\infty }^{2}\right) \right\} 
\overline{\rho }_{0}\cos \overline{R}\right.  \nonumber \\
&&\left. +\left\{ 2\left( \overline{R}^{2}-3\right) r_{0}^{2}r_{\text{opt}%
}^{2}-\kappa \overline{R}^{2}\left( 1-2v_{\infty }^{2}-\overline{R}%
^{2}\right) \right\} \overline{\rho }_{0}^{2}\cos \left( 2\overline{R}%
\right) \right.  \nonumber \\
&&\left. -\left\{ 36r_{0}^{2}r_{\text{opt}}^{2}+4\overline{R}^{2}r_{0}^{2}r_{%
\text{opt}}^{2}+6\kappa \overline{R}^{2}+2\kappa \overline{R}^{4}-12\kappa 
\overline{R}^{2}v_{\infty }^{2}\right\} \overline{\rho }_{0}\overline{R}\sin 
\overline{R}\right.  \nonumber \\
&&\left. +\left\{ 6r_{0}^{2}r_{\text{opt}}^{2}+\kappa \overline{R}%
^{2}-2\kappa \overline{R}^{2}v_{\infty }^{2}\right\} \overline{\rho }_{0}^{2}%
\overline{R}\sin \left( 2\overline{R}\right) \right] .
\end{eqnarray}

From the above expression, it is absurd to straightforwardly draw any
conclusion about stability or otherwise of the circular orbits. Clearly,
much will depend on the parameter ranges chosen on the basis of physical
considerations. While other parameters can be reasonably assigned, the as
yet unknown parameters are the dark matter radius $\kappa $ ($=2R_{\text{DM}%
}^{2}/\pi ^{2}$) and the dimensionless central density $\overline{\rho }_{0}$
($=8\rho _{0}R_{\text{DM}}^{2}/\pi $), again depending only on $\kappa $. In
the first order approximation, the density distribution in the dark matter
has been assumed in [4] to be low such that $8\pi G\kappa \rho
^{(0)}/c^{4}<<1$, but the central density $\rho _{0}$ could still be large
since $\left\vert \sin (x)/x\right\vert \leq 1$ [see Eq.(3)]. The question
therefore is how large or small could it be, or turning around, could there
be any upper limit on $\rho _{0}$ imposed by the stability criterion?

The answer to this question is yes and can be found graphically. We find
that $V^{\prime \prime }$ is indeed very sensitive to changes in $\rho _{0}$
leading to different upper limits $\rho _{0}^{\text{upper}}$ for different
galactic samples such that stable circular orbits are possible only when $%
\rho _{0}\leq \rho _{0}^{\text{upper}}$ in that sample. Different $\rho
_{0}^{\text{upper}}$ results from the fact that $R_{\text{DM}}$ changes from
sample to sample, as it should, and thereby leads to different \ (though not
too different) values for $\kappa $. For illustrative purposes, let us fix
typical values for a galactic sample, say UGC 0128, a low surface brightness
galaxy of moderate size, with the last observed scattering radius occurring
at $R_{\text{last}}=54.8$ kpc. Since direct observational data on $R_{\text{%
DM}}$ is yet unavailable for any sample, we approximately determine it by
projecting beyond $R_{\text{last}}$ the observed tendency of continuous
decline of the velocity dispersion to the zero value (see \textit{Fig.6 }of
Ref.[7]). For the present sample, one can then read off a value $R_{\text{DM}%
}\sim 88$ kpc, which corresponds to $\kappa =1.57\times 10^{3}$ kpc$%
^{2}=1.49\times 10^{46}$ cm$^{2}$ (conversion: $1$ kpc$=3.085\times 10^{21}$
cm). The other relevant parameters within the Newtonian approximation are%
\footnote{%
The range of $r$ and $r_{\text{opt}}$ is chosen so as to ensure $r/r_{\text{%
opt}}\gg r_{0}$. The usual formula for $r_{0}$ evaluates to $0.044$ for the
sample UGC 0128, while $r_{\text{opt}}\simeq 4R_{0}$. For this sample, $%
R_{0}=6.9$ kpc [7] such that $r_{\text{opt}}=27.6$ kpc. We emphasize that
exact values of these parameters including that of $\kappa $ are not
actually required as the behaviour of $V^{\prime \prime }$ \ is practically
insensitive to their variations within the Newtonian approximation.}: $%
r_{0}=0.044$, $v_{\infty }=0.000001$, $r_{\text{opt}}=27.6$ kpc, the
dimensionless radius $\overline{R}$ $(=\pi R/R_{\text{DM}})$ is chosen in
the range $\overline{R}$ $\in \lbrack 0.5\pi ,\pi ]$ and the dimensionless
density parameter in the range $\overline{\rho }_{0}\in \lbrack 0.25\pi ,$ $%
0.8\pi ]$.

Graphical analysis reveals a remarkable result: While $V^{\prime \prime }$
remains practically insensitive to the variation of the parameters ($\kappa
,r_{0}$, $v_{\infty }$, $r_{\text{opt}}$) within the periphery of Newtonian
approximation, \textit{it is greatly sensitive to the variation of the
remaining parameter }$\rho _{0}$. Figs.1 and 2 respectively show that, for
values of $\overline{\rho }_{0}>0.94$, there is instability in the entire or
partial range of the halo radii $\overline{R}$, while Fig.3 tells us that
there is an upper limit occurring at $\overline{\rho }_{0}^{\text{upper}}=$ $%
0.94=\lambda ^{\text{upper}}\pi $, where $\lambda ^{\text{upper}}=0.299$,
such that for $\overline{\rho }_{0}\leq \overline{\rho }_{0}^{\text{upper}}$
all circular orbits in the entire chosen radial range for $\overline{R}$ are
stable. Note that this value of $\lambda ^{\text{upper}}$ remains the 
\textit{same} under the change of parameter values ($\kappa ,r_{0}$, $%
v_{\infty }$, $r_{\text{opt}}$) in a given single sample, and so $\overline{%
\rho }_{0}^{\text{upper}}$ is quite robust for that sample.

Rewriting in terms of $\rho _{0}$, we have%
\begin{equation}
\rho _{0}^{\text{upper}}=\frac{\overline{\rho }_{0}^{\text{upper}}\pi }{8R_{%
\text{DM}}^{2}}=\frac{\lambda ^{\text{upper}}}{4\kappa },
\end{equation}%
which implies that $\kappa \rho _{0}^{\text{upper}}=$ constant, that is, the
larger the dark matter radius $R_{\text{DM}}$, the lower the value of $\rho
_{0}^{\text{upper}}$. This is an interesting feature of the EiBI model.
Plugging the values of $\lambda ^{\text{upper}}$ and $\kappa $, we find that
the constraint $\overline{\rho }_{0}\leq \overline{\rho }_{0}^{\text{upper}}$
immediately translates into an upper limit on $\rho _{0}$ such that for 
\begin{equation}
\rho _{0}\leq \rho _{0}^{\text{upper}}\text{,}
\end{equation}%
all circular orbits in the chosen range for $R$ are stable. We have verified
that $\overline{\rho }_{0}^{\text{upper}}$ keeps to the same value ($%
\overline{\rho }_{0}^{\text{upper}}=$ $0.94$) for many samples listed in
[6,7], the only difference to $\rho _{0}^{\text{upper}}$ thus comes from the
various $R_{\text{DM}}$ characteristic of various samples, as expressed in
Eq.(20). For UGC 0128, $\rho _{0}^{\text{upper}}\sim 9.9\times 10^{11}$ $%
M_{\odot }$kpc$^{-3}$, noting the conversion: $1$ cm$^{-2}=1.98\times 10^{59}
$ $M_{\odot }$kpc$^{-3}$. Thus the prediction is that, as long as $\rho _{0}$
of any galaxy obeys the stability induced constraint (21), the circular
material orbits in the halo around such galaxies will be stable up to a
maximum radius $R=R_{\text{DM}}$. Otherwise, the orbits will be unstable so
that there will be no dark matter around galaxies that have $\rho _{0}$
exceeding the $\rho _{0}^{\text{upper}}$. For further support to the
constraint, we may consider a large LSB sample U11748 with a projected dark
matter radius $R_{\text{DM}}\sim 106.18$ kpc, which leads to $\rho _{0}^{%
\text{upper}}=6.81\times 10^{11}$ $M_{\odot }$kpc$^{-3}$. This upper limit
is quite comparable with the values of other samples computed in this paper.
The NFW and Burkert profile fits yield $\rho _{0}=2.04\times 10^{8}$ $%
M_{\odot }$kpc$^{-3}$ and $1.67\times 10^{9}$ $M_{\odot }$kpc$^{-3}$
respectively [16], thus confirming (21).

To approximately quantify $\rho _{0}^{\text{upper}}$ in a familiar
situation, we consider our own Milky Way galaxy as an example. Note that
different groups have come up with somewhat different conclusions regarding
the local (solar neighbourhood) density of dark matter [8-10]. The density,
consistent with other standard estimates, seems to be $\rho _{0}^{\text{Solar%
}}=(0.3\pm 0.1)$ GeV/cm$^{3}=7.92\times 10^{6}$ $M_{\odot }$kpc$^{-3}$ [10].
Bergstrom, Ullio and Buckley [12] find local dark matter densities
acceptable in the range $0.2-0.8$ GeV/cm$^{3}$, so that the fitted values
roughly bunch around $\rho _{0}^{\text{Solar}}\sim 10^{6}$ $M_{\odot }$kpc$%
^{-3}$. For updated reviews, see [12,13]. With regard to the size of dark
matter around our galaxy, note that the observed velocity profile is
declining with radius: \textit{"The radial velocity dispersion shows an
almost constant value of }$120$\textit{\ km/s out to }$30$\textit{\ kpc and
then continuously declines down to }$50$\textit{\ km/s at about }$120$%
\textit{\ kpc"} [15]. Assuming a continuous fall, we can approximately take
the dark matter radius to be $R_{\text{DM}}\sim 180$ kpc. Eq.(20) then
yields a value $\rho _{0}^{\text{upper}}=2.37\times 10^{11}$ $M_{\odot }$kpc$%
^{-3}$. The actually fitted latest data on central density are $\rho _{0}^{%
\text{Milky Way}}=4.13\times 10^{7}$ $M_{\odot }$kpc$^{-3}$ (Burkert
profile) and $1.40\times 10^{7}$ $M_{\odot }$kpc$^{-3}$ (NFW profile) [16].
In both cases, we see that $\rho _{0}^{\text{upper}}$ is about four orders
of magnitude larger than the above fitted values of $\rho _{0}$, confirming
the inequality (21).

We should emphasize that it is the Doppler shifted light coming only from
material \textit{stable circular orbits} that led to the observation of
anomalous rotation velocities [19]. That is precisely the reason why we
focused on the circular orbits obtaining a stability inspired constraint. It
is possible however that non-circular orbits or different kinds of
instabilities may lead to stronger constraints than the ones derived in this
work. For example it could be that galaxies with stable circular orbits,
such as the Milky Way, actually present other unstable orbits and thus
becomes non-viable within the EiBI scenario.\footnote{%
We thank an anonymous reviewer for pointing out this possibility.} All kinds
of orbits are of course dictated by the potential and in the present case
the exact form, from Eq.(12), is%
\begin{eqnarray}
V(\overline{r}) &=&\frac{(\overline{\rho }_{0}\sin \overline{r}-\overline{r}%
)(\overline{r}-\overline{\rho }_{0}\sin \overline{r}+\overline{\rho }_{0}%
\overline{r}\cos \overline{r})}{\omega \overline{r}^{2}}-L^{2}\left[ \frac{%
\overline{r}-\overline{\rho }_{0}\sin \overline{r}+\overline{\rho }_{0}%
\overline{r}\cos \overline{r}}{\omega \overline{r}^{3}}\right]   \nonumber \\
&&+E^{2}\left[ -1+\frac{\left( q\overline{r}^{2}+r_{0}^{2}\right)
^{v_{\infty }^{2}}(\overline{r}-\overline{\rho }_{0}\sin \overline{r})(%
\overline{r}-\overline{\rho }_{0}\sin \overline{r}+\overline{\rho }_{0}%
\overline{r}\cos \overline{r})}{\omega \overline{r}^{2}}\right] 
\end{eqnarray}%
where $q=\left( R_{\text{DM}}/r_{\text{opt}}\right) ^{2}$, $\omega =\left(
R_{\text{DM}}/\pi \right) ^{2}$, $\overline{\rho }_{0}$ and $v_{\infty }^{2}$
are the galactic sample characteristics, while $E^{2}$ and $L^{2}$ are
constants describing arbitrary particle trajectories. Once a sample is
chosen, one can specify ranges for $E^{2}$ and $L^{2}$ to see if the
potential allows turning points so that non-circular orbits can lie between
them (see, e.g., [20]). From the above, it is seen that the potential $V(%
\overline{r})$ is a combination of periodic ($\sin \overline{r}$)\ and
aperiodic ($\overline{r}\cos \overline{r}$) functions. It turns out that,
depending on the sample as well as on the prescription of $E^{2}$ and $L^{2}$%
, the spacetime may or may not support turning points. Also, Eq.(11) is
non-integrable for this complicated potential so that orbit equations in
closed form cannot be obtained. We believe that definitive conclusions about
the possibility of other (un)stable orbits can nonetheless be drawn by
adopting the more powerful method of autonomous dynamical system for Eq.(11)
with this potential. That would by itself be a separate task, which we leave
for the future.

In conclusion, we can say that the stability of orbits is a powerful
constraint governing the dynamics of particles in the galactic halo (see
e.g., [2,17-19]). The obtained stability inspired constraint (21) is a 
\textit{new} testable prediction of the EiBI theory of dark matter developed
by Harko \textit{et al.} [4]. The theory is increasingly gaining importance
after it was originally developed by Ba\~{n}ados and Ferreira [21] on the
basis of modified matter-gravity coupling [22-26]. It is powerful enough to
handle different physical scenarios such as neutron stars and other compact
objects first studied in Refs.[27-32]. Consistency with well known density
profiles, as shown above, indicates that the EiBI theory could provide yet
another viable alternative model for dark matter. However, a solution beyond
the current Newtonian approximation and conclusively observed data on $R_{%
\text{DM}}$ are the two ingredients that will be needed for a more precise
testing of the theory in future. Also, note that if for just one galaxy, or
cluster, the central dark matter density is higher than the predicted value $%
\rho _{0}^{\text{upper}}\sim 10^{12}$ $M_{\odot }$kpc$^{-3}$, then the EiBI
scenario would fail as a candidate for explaining the anomalous rotation
velocity of galaxies.

\bigskip 

\textbf{Acknowlegments}

One of us (Ramil Izmailov) was supported by the Ministry of Education and
Science of Russian Federation. This work was supported in part by an
internal grant of M. Akmullah Bashkir State Pedagogical University in
the field of natural sciences. The authors are thankful to Guzel Kutdusova,
Regina Lukmanova and Almir Yanbekov for technical assistance.

\newpage
\begin{center}
\textbf{Figure captions}
\end{center}

\begin{figure} [h]
\includegraphics [width=\linewidth] {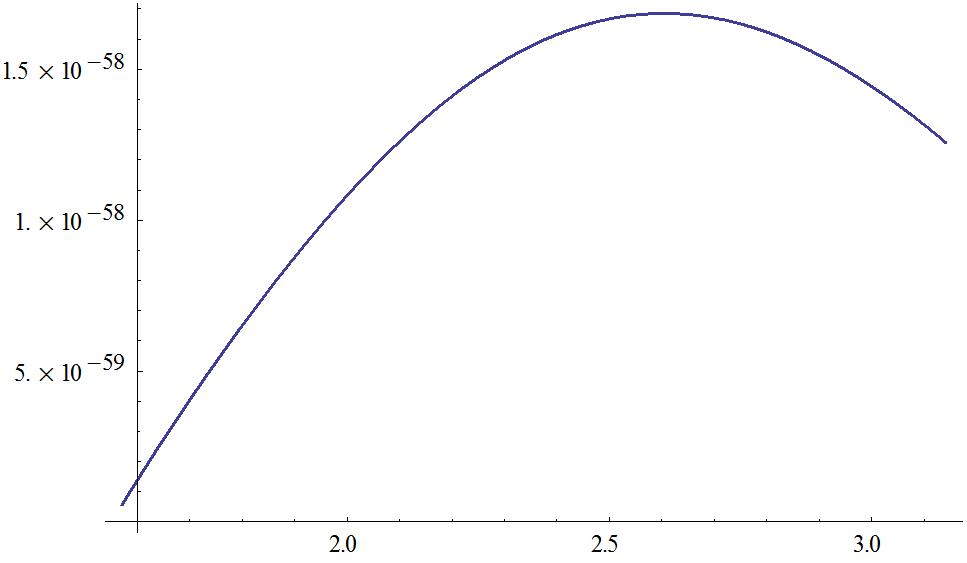}
\caption{Plot of $V^{\prime \prime }$ vs $\overline{R}$ specific to the sample
UGC 0128. The chosen parameters are: $r_{0}=0.044$, $v_{\infty }=0.000001$, $%
r_{\text{opt}}=27.6$ kpc, $\kappa =1.57\times 10^{3}$ kpc$^{2}$ and $%
\overline{\rho }_{0}=0.50\pi $. The orbits are unstable in the chosen entire
radial range $\overline{R}$ $\in \lbrack 0.5\pi ,\pi ]$ because $V^{\prime
\prime }>0$ there}
\end{figure}

\begin{figure} 
\includegraphics [width=\linewidth] {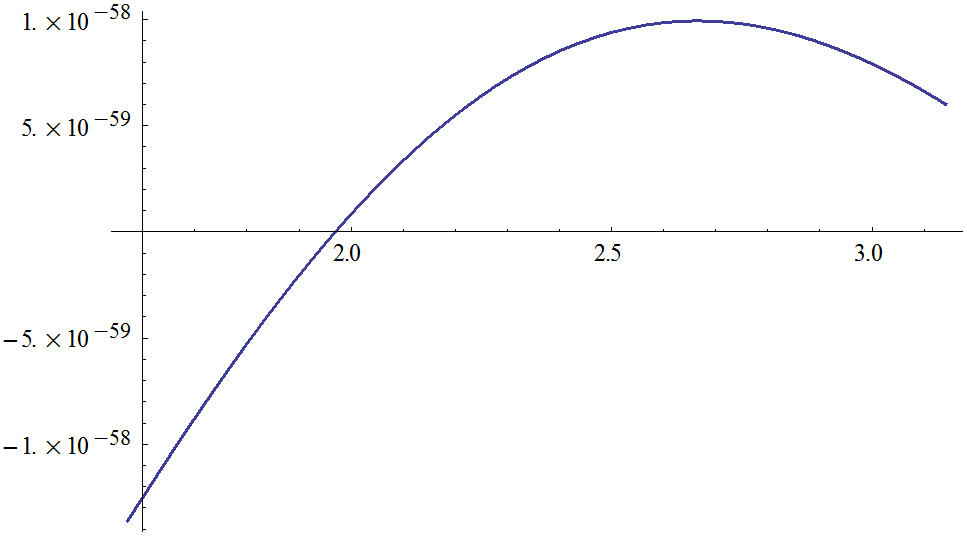}
\caption{Plot of $V^{\prime \prime }$ vs $\overline{R}$ specific to the sample
UGC 0128 The chosen parameters are: $r_{0}=0.044$, $v_{\infty }=0.000001$, $%
r_{\text{opt}}=27.6$ kpc and $\kappa =1.57\times 10^{3}$ kpc$^{2}$. Here
central density is further lowered to $\overline{\rho }_{0}=0.34\pi $. The
orbit is unstable in some intermediate radii as $V^{\prime \prime }$ is
partly positive and partly negative}
\end{figure}

\begin{figure}
\includegraphics [width=\linewidth] {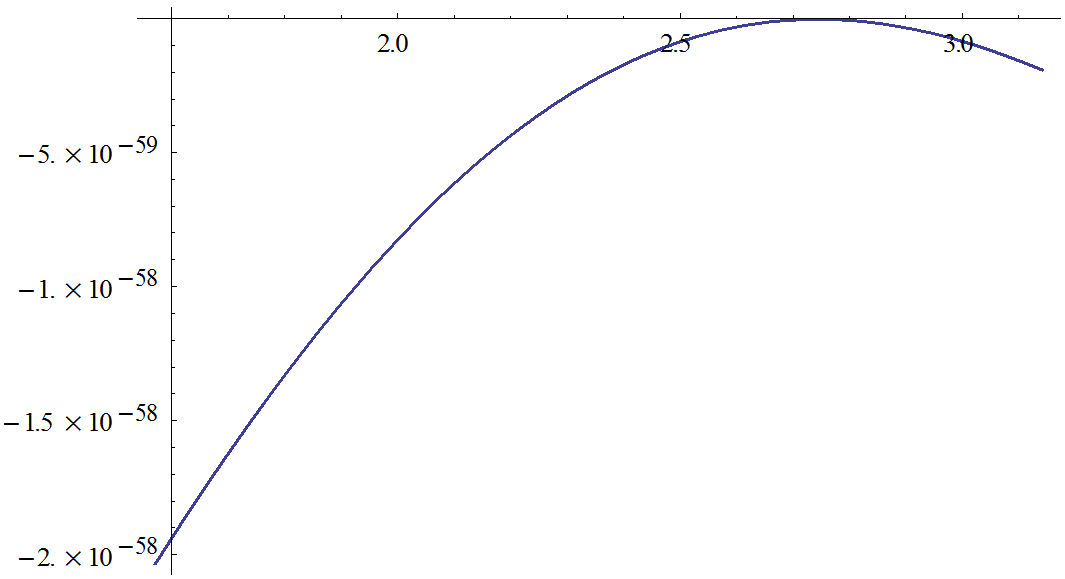}
\caption{Plot of $V^{\prime \prime }$ vs $\overline{R}$ specific to the sample
UGC 0128. The chosen parameters are: $r_{0}=0.044$, $v_{\infty }=0.000001$, $%
r_{\text{opt}}=27.6$ kpc and $\kappa =1.57\times 10^{3}$ kpc$^{2}$. Here
central density is further lowered to $\overline{\rho }_{0}=0.299\pi $,
which corresponds to $\overline{\rho }_{0}^{\text{upper}}=$ $0.94=\beta ^{%
\text{upper}}\pi $, where $\beta ^{\text{upper}}=0.299$. The orbit is 
\textit{stable} in the entire chosen range for $\overline{R}$ $\in \lbrack
0.5\pi ,\pi ]$.}
\end{figure}

\newpage

\textbf{REFERENCES}

[1] M.K. Mak and T.Harko, Phys. Rev. D \textbf{70}, 024010 (2004); 
T.Matos, F.S. Guzm\'{a}n and D. Nu\~{n}ez, Phys. Rev. D \textbf{62}, 061301(2000); 
P.J.E. Peebles, Phys. Rev. D \textbf{62}, 023502 (2000);  
U.Nucamendi, M. Salgado and D. Sudarsky, Phys. Rev. D \textbf{63}, 125016(2001); 
E.W. Mielke and F.E. Schunck, Phys. Rev. D \textbf{66}, 23503(2002); 
S. Capozziello, M. De Laurentis and S. D. Odintsov,  Mod. Phys. Lett. A \textbf{29}, 1450164 (2014); 
C.-Q. Geng, D. Huang, and L.-H. Tsai, Mod. Phys. Lett. A \textbf{29}, 1440003 (2014);
S. Bharadwaj and S. Kar, Phys. Rev. D \textbf{68}, 023516 (2003); 
M.Colpi, S.L. Shapiro and I. Wasserman, Phys. Rev. Lett., \textbf{57}, 2485 (1986); 
A.C. Pope \textit{et al.}, Astrophys. J. \textbf{607}, 655 (2004); 
M. Tegmark et al., Phys. Rev. D \textbf{69}, 103501 (2004a); 
M.Tegmark et al., Astrophys. J. \textbf{606}, 702 (2004b); 

[2] F. Rahaman, K.K. Nandi, A. Bhadra, A. Kalam and K. Chakraborty, Phys.
Lett. B \textbf{694}, 10 (2010)

[3] T. Harko and F.S. N. Lobo, Phys. Rev. D\textbf{\ 83}, 124051 (2011)

[4] T. Harko, F.S.N. Lobo, M.K. Mak and S.V. Sushkov, Mod. Phys. Lett. A 
\textbf{29}, 1450049 (2014)

[5] P. Salucci and M. Persic, \textit{in Dark and visible matter in galaxies}%
, eds. M. Persic and P. Salucci, ASP Conference Series \textbf{117}, 1 (1997)

[6] P.D. Mannheim and J.G. O'Brien, Phys. Rev. Lett. \textbf{106}, 121101
(2011)

[7] P.D. Mannheim and J.G. O'Brien, arXiv:1011.3495 (2011)

[8] J. N. Bahcall, M. Schmidt and R. M. Soneira, Astrophys. J. \textbf{265},
730 (1983)

[9] R. R. Caldwell and J. P. Ostriker, Astrophys. J. \textbf{251}, 61 (1981)

[10] M. S. Turner and F. Wilczek, Phys. Rev. D \textbf{42}, 1001 (1990)

[11] J. Bovy and S. Tremaine, Astrophys. J.\textbf{\ 756}, 89 (2012)

[12] L. Bergstrom, P. Ullio and J. H. Buckley, Astropart. Phys. \textbf{9},
137 (1998)

[13] G. Bertone, D. Hooper and J. Silk, Phys. Rep. \textbf{405}, 279 (2005)

[14] J.I. Read, J. Phys. G: Nucl. Part. Phys.\textbf{\ 41}, 063101 (2014)

[15] G. Battaglia \textit{et al.}, Mon. Not. R. Astron. Soc. \textbf{364},
433 (2005); Erratum: \textbf{370}, 1055B (2006)

[16] F. Nesti and P. Salucci, JCAP \textbf{07 }(2013) 016

[17] K.K. Nandi \textit{et al.}, Mon. Not. R. Astron. Soc. \textbf{399},
2079 (2009)

[18] K.K. Nandi and A. Bhadra, Phys. Rev. Lett. \textbf{109}, 079001 (2012)

[19] K. Lake, Phys. Rev. Lett. \textbf{92}, 051101 (2004)

[20] J.B.Hartle, \textit{Gravity: An Introduction to Einstein's General
Relativity}, Pearson Education Inc. (2003), p.221

[21] M. Ba\~{n}ados and P. G. Ferreira, Phys. Rev. Lett. \textbf{105},
011101 (2010)

[22] S. Nojiri and S.D. Odintsov, Phys. Lett. B \textbf{599}, 137 (2004)

[23] S. Nojiri and S.D. Odintsov, Phys. Rep. \textbf{505}, 59 (2011)

[24] O. Bertolami, C. G. Boehmer, T. Harko and F. S. N. Lobo, Phys. Rev. D 
\textbf{75}, 104016 (2007)

[25] T. Harko and F. S. N. Lobo, Eur. Phys. J. C \textbf{70}, 373 (2010)

[26] T. Delsate and J. Steinhoff, Phys. Rev. Lett. \textbf{109}, 021101
(2012)

[27] P. Pani and V. Cardoso, Phys. Rev. Lett. \textbf{107}, 031101 (2011)

[28] P. Pani, T. Delsate and V. Cardoso, Phys. Rev. D\textbf{\ 85}, 084020
(2012)

[29] V.H. Robles and T. Matos, Mon. Not. R. Astron. Soc. \textbf{422}, 282\
(2012)

[30] I. L. J. Casanellas, P. Pani, and V. Cardoso, Astrophys. J. \textbf{745}, 15 (2012)

[31] P. Pani, V. Cardoso and T. Delsate, Phys. Rev. Lett. \textbf{107},
031101 (2011)

[32] T. Harko, F.S.N. Lobo, M.K. Mak and S.V. Sushkov, Phys. Rev. D \textbf{88}, 044032 (2013)

\bigskip

\end{document}